\documentclass{elsart}

\usepackage{graphicx}
\usepackage[square,comma,numbers,sectionbib,sort&compress]{natbib}
\usepackage{bm}                 
\usepackage{amsmath}            
\usepackage{amsfonts}
\usepackage{amssymb}

\begin{document}
\begin{frontmatter}

\title{Entanglement and its Role in Shor's Algorithm}

\author[leeds,qols]{Vivien M.~Kendon},
\ead{V.Kendon@leeds.ac.uk}
\author[hp,qols]{William J.~Munro}

\address[leeds]{School of Physics and Astronomy, University of Leeds, LS2 9JT, UK}
\address[hp]{Hewlett-Packard Laboratories, Filton Road, Stoke Gifford, Bristol, BS34 8QZ, UK}
\address[qols]{QOLS, Optics, Blackett Laboratory, Imperial College London, SW7 2BW, UK}

\begin{abstract}
Entanglement has been termed a critical resource for quantum 
information processing and is thought to be the reason 
that certain quantum algorithms, such as Shor's factoring algorithm, can 
achieve exponentially better performance than their classical counterparts. 
The nature of this resource is still not fully understood: here  
we use numerical simulation to
investigate how entanglement between register qubits
varies as Shor's algorithm is run on a quantum computer.
The shifting patterns in the entanglement are found to 
relate to the choice of basis for the quantum Fourier transform.
\end{abstract}


\end{frontmatter}

\section{Introduction}
\label{sec:intro}

Quantum computation has the potential to provide
significantly more powerful computers than classical computation
-- if we can build them.
There are numerous possible routes 
forward for quantum hardware \cite{colls00}, however, progress in the 
development of algorithms has been slow, in part because we
don't yet fully understand how the quantum advantage works.
There are two key characteristics of the quantum resources
used for computation.
The first is that a general superposition 
of $2^n$ levels may be represented in $n$ 2-level systems 
\cite{jozsa98a}, allowing the 
the physical resource to grow only {\it linearly} with $n$
(quantum parallelism).
The second aspect is best explained by
considering the classical computational cost of
simulating a typical step in a quantum computation.
If entanglement is absent then the algorithm can be 
simulated with an equivalent amount of classical resources.
Jozsa and Linden \cite{jozsa02a} showed that,
if a quantum algorithm cannot be simulated classically using resources 
only polynomial in the size of the input data, then it will have multipartite 
entanglement involving unboundedly many of  its qubits -- if it is run 
on a quantum computer using pure quantum states.
However, the presence of multipartite entanglement 
is not a sufficient condition for a pure state quantum computer to 
be hard to simulate classically.  As Jozsa and Linden point out,
if the quantum computer is described
using stabilizer formalism \cite{nielsenchuang00,gottesman97a},
there are many highly entangled states that have simple 
classical descriptions.  Moreover, a quantum computer
using mixed states may still require exponential classical resources
to simulate even if its qubits are not entangled, and it is not known whether
such states may be used to perform efficient computation.
In any case, being hard to simulate classically doesn't imply
the quantum process is doing any useful computation.
If we want to understand quantum computation, we will have to
look more closely at specific examples.

Few quantum algorithms provide an exponential speed up over classical
algorithms, of those that do, Shor's algorithm (order-finding) \cite{shor95a}
is perhaps the most important because it can 
be used to factor large numbers and hence has 
implications for classical security methods.
There is no proof that an equally efficient classical
algorithm cannot exist for Shor's algorithm, and it is worth noting
that a sub-exponential algorithm has been found recently for the related problem
of primality testing \cite{agrawal04a}.
Proving a speed up is in general a tough task, few such proofs exist
for exponential speed up of quantum over classical, one example being
a quantum walk algorithm with a proven exponential speed
up (w.r.t.~an oracle) \cite{childs02a}.

Assuming that Shor's algorithm does provide an exponential speed up, and
given that multipartite entanglement is necessary (though not sufficient)
for pure state quantum computation with an exponential speed up over
classical computation, in this paper we investigate what the entanglement
is doing during the computational process, as it proceeds, gate by gate.
To be clear, we reiterate that we are not trying to prove 
\textit{whether} entanglement is present, we take it as given that
there will be at least $\log r$ of entanglement entropy (where $r$ is the
period being determined, and logs are in base 2 throughout this paper)
at the mid-point of Shor's algorithm,
as first shown by Nielsen and Chuang \cite{nielsen95a}.
Instead, we would like to know what role it plays in the computation.
What we have in mind is a role comparable to the role of entanglement in
quantum communications, where a maximally entangled pair of qubits
can be used to perform specific communications tasks (such as
teleportation of an unknown quantum state \cite{bennett93a},
or transmission of two classical bits of information \cite{bennett92a}),
which consume the entanglement in direct
proportion to the amount of communication achieved.
To date, little has been said about what role entanglement actually
plays in quantum computation, the focus has been almost entirely on
proving it is present, in sufficient quantities to make classical simulation
inefficient (besides \cite{jozsa02a}, see, for example,
\cite{vidal03a,orus03a,ukena05a,shimoni05a}).
We aim to throw some light on the question of
what function it plays by calculating the entanglement as it varies during
the course of the execution of Shor's algorithm, and looking at how
it correlates with the progress of the algorithm.

For this study, we are using a standard gate sequence for Shor's 
algorithm using pure quantum states.
Use of mixed states and different gate sequences may produce different
entanglement patterns, but if there is a crucial role for entanglement
in the computation, it will be identifiable as a common feature of all
implementations.
Parker and Plenio \cite{parker01a} have presented a version of Shor's
algorithm using only one pure qubit, the rest may start in any mixed state.
They confirmed (numerically) that entanglement was present when the algorithm
ran efficiently for factoring 15 and 21.

We should also add that, since we are investigating the logical functioning of
the algorithm, we are not concerned with any practical questions of
imperfect gates, decoherence, etc., nor with optimising the gate sequences
given constraints on the number of qubits or the types of gates
available.  Much valuable work has been done in these areas by many authors,
notably Vedral et al, \cite{vedral96a},
Gossett \cite{gossett98a}, and
Van Meter and Itoh \cite{vanmeter05a}
on constructing efficient operations from elementary gates;
Zalka \cite{zalka98a} and
Beauregard \cite{beauregard03a} who optimise the overall operation of
Shor's algorithm in fewer qubits; and
Fowler and Hollenberg \cite{fowler04a} who analyse scalability and
accuracy. 

The organisation of this paper is as follows.  We start with a brief overview
of Shor's algorithm in \S\ref{sec:shor}, to set up our notation.  This 
is followed in \S\ref{sec:shor15} by a discussion of how the entanglement
varies in an instance of factoring 15, which also serves to introduce
the entanglement measures we are using.  In \S\ref{sec:entangle21}
the pattern of entanglement in several examples of factoring 21 is presented.
Larger semiprimes are tackled in \S\ref{sec:entanglement}, from which we
are able to deduce our main conclusions,
which are summarised in \S\ref{sec:discussion}.

\section{Shor's Algorithm}
\label{sec:shor}

We begin with a brief overview of how Shor's algorithm works,
in order to remind the reader and to establish our notation.
We wish to factor a number $N=pq$ 
where $p$ and $q$ are prime numbers. 
Classical number theory provides a way to determine these primes
with high probability (not unity generally) by finding the period $r$
of the function 
\begin{equation}
f_x (a)=x^a (\text{mod $N$}),
\end{equation}
where $x$ is an integer chosen to be less than $N$ and co-prime to it, and
$a \in \mathbb{Z}$.
It is efficient to check whether $x$ is co-prime to $N$
using Euclid's algorithm \cite{knuth81a}.  If $x$ happens not to be
co-prime then their common factor gives a factor of $N$ and the job is done,
but this happens only rarely for large $N$.
Once the period $r$ is found, the numbers 
\begin{equation}
m_{\pm}=x^{r/2} \pm 1
\label{eq:xr}
\end{equation}
generally share either $p$ or $q$ with $N$ as a common factor.
Not all choices of $x$ give periods $r$ which yields a factor $p$ or $q$.
For instance, sometimes the period $r$ will be odd, whence the numbers from 
eq.~(\ref{eq:xr}) can be non-integer.
When the chosen $x$ does not lead to a valid factor,
the procedure can be repeated with a different
choice until a factor is found.  This is efficient since
the probability of success is at least $\frac{1}{2}$ per trial for
the case of semiprimes (see Shor \cite{shor95a}).

\begin{figure}[!tb]
    \begin{center}
	\includegraphics[scale=0.33]{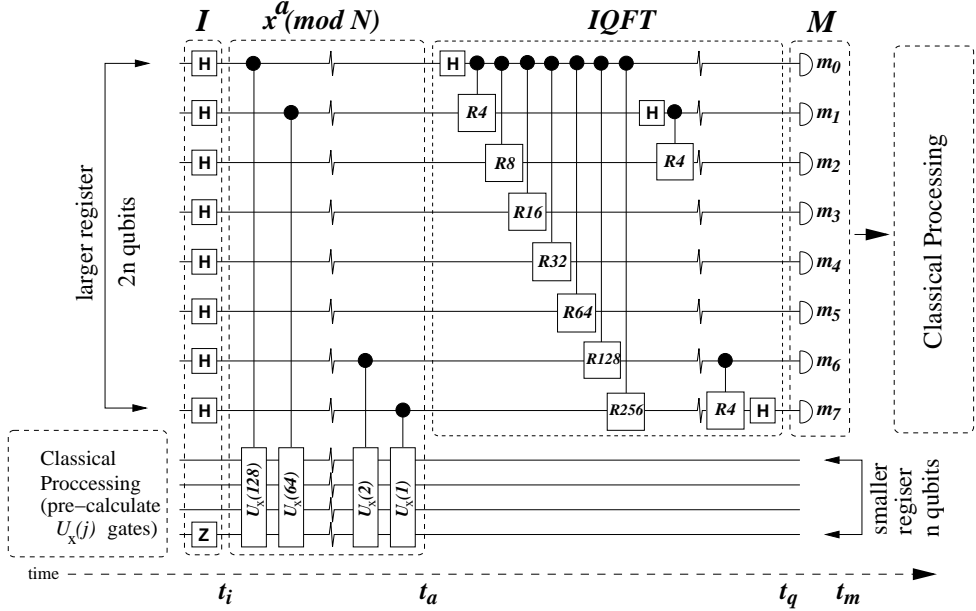}
    \end{center}
    \caption{Schematic circuit diagram of Shor's algorithm for factoring 15
	implemented on a 12 qubit quantum register.
	The initialisation $I$ is done with single qubit Hadamard (H) and
	bit-flip (Z) gates.  Controlled-$U_x(j)$ gates are used to
	produce $x^a (\text{mod $N$})$.
	The inverse quantum Fourier transform (IQFT)
	uses controlled rotations ($Rm$).
	The last quantum step is the measurement (M), which is followed by
	classical post-processing to obtain a factor of $N$.}
    \label{fig:15gates}
\end{figure}

The hard part of the algorithm 
is determining the period $r$ of the function 
$f_x (a)=x^a (\text{mod $N$})$. Shor found a very elegant and 
efficient means of doing this quantum 
mechanically, depicted schematically in fig.~\ref{fig:15gates}. 
Consider that one has two quantum registers 
(one of size $2n$ where $n=\lceil\log N\rceil$ qubits and the second 
of size $n$ qubits.
We will denote the basis states of a quantum register by 
$|a\rangle$, with $a\in\{0\dots 2^{2n}-1\}$.  The binary representation 
of $a$ indicates which register qubits are in the state representing zero and
which are in the state representing one.
A general state of a $2n$ qubit register
$|\psi(t)\rangle$ at time $t$
can thus be written as a superposition of basis states,
\begin{equation}
|\psi(t)\rangle=\sum_{a=0}^{2^{2n}-1}\alpha_a(t)|a\rangle,
\label{eq:rstate}
\end{equation}
where $\alpha_a(t)$ is a complex number, normalised such that
$\sum_{a=0}^{2^{2n} -1} |\alpha_a(t)|^2=1$.
The algorithm begins by preparing the larger quantum register in 
an equal superposition  $(2^{2n})^{-1/2}\sum_{a=0}^{2^{2n} -1} |a\rangle$ 
of all possible $2^{2n}$ basis states while the 
smaller register is prepared in the definite state $|1\rangle$ ($\equiv
|0\dots01\rangle$). 
The initial state of both registers is thus
\begin{eqnarray}
|\Psi(t_i)\rangle = \frac{1}{2^n} \sum_{a=0}^{2^{2n} -1} |a\rangle |1\rangle
\label{eq:istate}
\end{eqnarray}
The next step is a unitary transformation which acts on both registers 
according to $U |a\rangle |b\rangle = |a\rangle |b x^a (\text{mod $N$})\rangle$ giving 
the output state
\begin{eqnarray}\label{eqn-U}
|\Psi(t_a)\rangle= \frac{1}{2^n} \sum_{a=0}^{2^{2n} -1} |a\rangle |x^a (\text{mod $N$})\rangle
\end{eqnarray}
Then an inverse quantum Fourier transform (IQFT) defined by
\begin{eqnarray}
Q^{-1} |y\rangle = \frac{1}{2^n} \sum_{z=0}^{2^{2n} -1} \e^{-2\pi i yz/2^{2n}}|z\rangle 
\end{eqnarray}
is applied, which transform the state $|\Psi(t_a)\rangle$
from eq.~(\ref{eqn-U}) into
\begin{eqnarray}
|\Psi(t_q)\rangle=
 \frac{1}{2^{2n}} \sum_{a=0}^{2^{2n} -1} \sum_{z=0}^{2^{2n} -1} \e^{-2\pi i az/2^{2n}} 
 |z\rangle |x^a (\text{mod $N$})\rangle.
\end{eqnarray}
By measuring the larger register in the computational basis we obtain 
an integer number $c$.  Now $c/2^{2n}$ is a close approximation
to the fraction $k/r$, where $0\le k < r$, the value of $k$
varies depending on the particular value of $c$ that is measured,
but both $r$ and $k$ can be obtained classically using continued fractions
(provided $c \ne 0$).  Choosing the larger register to be $2 n$ qubits
provides a high enough accuracy for $c$ such that $r$ can be 
determined from a single measurement on all $2 n$ qubits. It is possible 
to use fewer qubits in this first register but the probability of 
correctly determining $r$ decreases, and the algorithm may need to be repeated 
correspondingly many more times.  If $r$ is not prime, and happens to share
a factor $p$ with $k$, then one also fails to determine $r$ correctly,
instead obtaining $r/p$.  Again, this only reduces the probability of
success by a factor polynomial in $N$, so the exponential nature of the
speed up is maintained.

\section{Factoring 15}
\label{sec:shor15}

\begin{figure}[!tb]
    \begin{center}
	\includegraphics[scale=0.33]{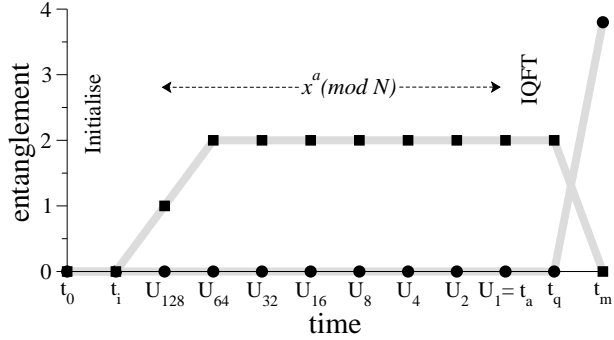}
    \end{center}
    \caption{Entanglement between the two registers (squares) and
	within the smaller register (circles) 
	in Shor's 12 qubit algorithm as a function of gates 
	sequence according to fig (\ref{fig:15gates}) with the co-prime
	chosen as $x=13$.
	The entanglement within the larger register is zero throughout.}
    \label{fig:15-13ent}
\end{figure}
We start our analysis of the entanglement by studying the circuit for
factoring 15 (3x5), though 
it is not necessarily typical of factoring larger numbers.
Since many gates make no change to the entanglement,
rather than tracking the entanglement as each basic gate is applied,
we choose to look at certain key points in the algorithm.
We restrict our attention to controlled composite gates:
the $U_x(j)$ gate which implements the operation 
$x^j (\text{mod $N$})$ for $j\in\{1,2,4\dots 2^{2n}\}$,
and the rotations in the IQFT.
Details of how to 
efficiently construct these composite gates from a universal set of one 
and two qubit gates may be found in, for example, \cite{nielsenchuang00}. 
There are 8 of the $U_x(j)$ gates (in general $2n$, one for each
larger register qubit), which is manageable, but for the IQFT
there are 27 (in general $(2n+1)(n-1)$ for a $2n$ qubit register)
rotation gates: for our purposes in this paper it is sufficient to treat
the whole IQFT as one unit.
Along with single qubit gates as necessary, the circuit 
using these composite gates is depicted in fig.~\ref{fig:15gates}.

As we are only considering the evolution of pure states we can
measure the entanglement between the two registers
using the entropy of the subsystems
\begin{equation}
E_c = -\sum_i \lambda_i\log\lambda_i,
\label{eq:Esubsys}
\end{equation}
where the $\{\lambda_i\}$ are the eigenvalues of the reduced density
matrix of either of the registers (both have the same eigenvalues).
The reduced density matrix of one of the registers
is obtained from the full pure state of
the system by applying a partial trace over the other register,
\begin{equation}
\rho_L(t) = \text{Tr}_S |\Psi(t)\rangle\langle\Psi(t)|,
\end{equation}
and similarly $\rho_S(t) = \text{Tr}_L |\Psi(t)\rangle\langle\Psi(t)|$,
where $L$ and $S$ denote the larger and smaller registers respectively.

To quantify the entanglement \emph{within} each register is not so 
straightforward.
Most entanglement measures for mixed states, such as $\rho_L$ and $\rho_S$,
are computationally intractable in practice for more than a few qubits;
we also need to consider all the possible divisions of the qubits
into different subsets in order to locate all of the entanglement.
A reasonable approximation to quantifying the entanglement within a
register can be obtained by applying a partial transpose to each possible subset
of qubits and calculating the negativity \cite{peres96a,zyczkowski98a}
given by $\eta= \text{Tr} |\rho^{T}| - 1$  i.e., the sum of
the negative eigenvalues of the transposed matrix $\rho_L^T$ or $\rho_S^T$.
If the negativity is zero for all possible subsets of qubits in the register,
then we can say that at most the register has bound entanglement
\cite{horodecky98a}, which
is not generally considered useful for quantum information tasks
(though see \cite{horodecky98b}).
Non-zero negativity for any subset of qubits
definitely indicates the presence of entanglement (across that particular
division).

Finally, we use the entanglement of formation \cite{wootters97a} to
quantify the pairwise entanglement between two qubits.
The entanglement of formation quantifies how much entanglement is
needed to make the state from unentangled ingredients.  In general
it is hard to calculate explicitly, however, Wootters
\cite{wootters97a} found an elegant formula for the case of two
qubits in a mixed state $\rho$, the concurrence $C$ is given by
$C = \text{max}(\lambda_1 - \lambda_2 - \lambda_3 - \lambda_4, 0)$,
where the $\lambda_i$ are the square roots of the eigenvalues of
$\rho \tilde{\rho} = \rho\;\sigma_{y}^{A}\!\otimes\!\sigma_{y}^{B}
\rho^{*} \sigma_y^A\!\otimes\!\sigma_{y}^{B}$,
with $A$, $B$ labels for the two qubits, $\sigma_y$ the Pauli spin operator,
and $\rho^{*}$ denotes the complex conjugation of $\rho$
in the computational basis $\{|00\rangle, |01\rangle,|10\rangle,|11\rangle\}$.
The entanglement of formation 
$E_f = h(\frac{1}{2} + \frac{1}{2}\sqrt{1-C^2})$,
where $h(\eta)$ is the binary entropy function,
$h(\eta) = -\eta\log(\eta) - (1-\eta) \log(1-\eta)$.
We use $E_f$ to check for pairwise entanglement within and between
the qubit registers.
Note that quantum states can be highly entangled even
without containing any pairwise entanglement \cite{coffman99a,kendon02a}.

In fig.~\ref{fig:15-13ent} we plot the entanglement in Shor's algorithm
using the entropy of the subsystem where possible (full state is pure)
and the negativity where the single register state is mixed.
The negativity turns out to be zero for both registers throughout the
algorithm (except the measurement leaves the smaller register 
entangled, but this cannot be useful for the remaining
classical steps of the algorithm).
The entanglement between the registers builds up to a maximum during
the first two $U_x(j)$ gates, then stays constant until the measurement.
We also note (from calculating the entanglement of formation for 
appropriate pairs of qubits)
that there is no pairwise entanglement between any pair of qubits
at any of the sampled points in this instance of the algorithm,
neither within nor between the registers.

Since the IQFT is the crucial step for finding the period, we
looked in more detail at how the distribution of the
entanglement changes over the operation of the IQFT.
\begin{figure}[!tb]
    \begin{center}
	\includegraphics[scale=0.33]{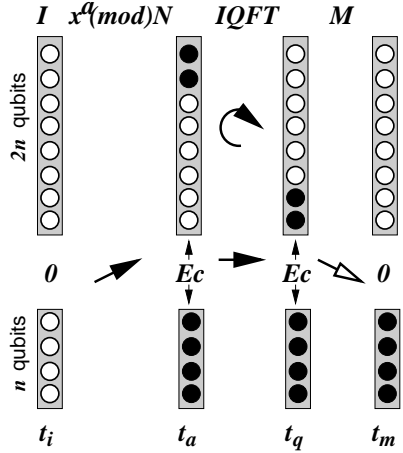}
    \end{center}
    \caption{Pattern of entanglement during Shor's algorithm factoring
	$N=15$ with co-prime $x=13$.
	After the $U_x(j)$ gates the top two qubits in the larger register
	(filled) are entangled with the four qubits in the smaller register.
	After the IQFT, the entanglement is transfered to the lower two
	qubits in the larger register. Qubits represented by open circles
	are not entangled. Time sequence corresponds to fig.~\ref{fig:15gates}.}
    \label{fig:15pat}
\end{figure}
Unitary operations can only alter entanglement within the elements they
are applied to.   Applying this principle to the circuit in
fig.~\ref{fig:15gates},
since the entanglement within each register is zero (strictly speaking, zero
apart from possible bound entanglement) after the
modular exponentiation, it is clear that entanglement cannot
decrease during the IQFT, since no further gates act where the only significant
entanglement is located, between both registers.
Furthermore, since each pair of qubits in the upper
register has an entangling (2-qubit) gate applied to it only once
during the algorithm, entanglement within the upper register can
only be generated or shifted around, not decreased.
And indeed our numerical calculations show 
the distribution of entanglement between the individual qubits does change
in our example of factoring 15 with $x=13$.
By examining the entropy of each possible subset of qubits in each
register, we can deduce that only two of the eight
qubits are entangled with the four qubits in the smaller register, and 
during the action of the IQFT, this entanglement is transfered
from the top two qubits to the bottom two in the larger register.
We represent this schematically in fig.~\ref{fig:15pat}.

However, we should remember that 15 is actually extremely easy 
to factor. It is straightforward to see that at least one of $x^{r/2}\pm 1$ 
is divisible by 3 or 5 for almost any random choice of $x,r >1$, regardless of 
whether $x$ is co-prime to $N$ or whether $r$ is the period of $x^a 
(\text{mod $N$})$.  To learn anything significant,
we need to look at more examples.

\section{Factoring 21}
\label{sec:entangle21}

\begin{table}[!bt]
\caption{Average entropy of subsystems for factoring 21 with $x=2$, and
average negativity (after the IQFT) for different sized subsystems of the
larger register.}
\label{tab:21-2}
\begin{center}
\begin{tabular*}{\textwidth}{@{\extracolsep{\fill}}c@{}c@{}c@{}c@{}c@{}c@{}}
\hline
size of & & \multicolumn{2}{c}{large register} & large register & large register\\
subsystem & small register & after $U$ & after IQFT & difference $\Delta E_1$ & negativity \\
\hline
 1 qubit    &   0.811   &  1.000  &   0.938  &   -0.062 & 0.172 \\
 2 qubits   &   1.538   &  1.600  &   1.599  &   -0.001 & 0.397 \\
 3 qubits   &   2.151   &  1.843  &   2.020  &   +0.177 & 0.591 \\
 4 qubits   &   2.585   &  1.972  &   2.283  &   +0.311 & 0.678 \\
 5 qubits   &   2.585   &  2.081  &   2.447  &   +0.366 & 0.749 \\
\hline
 6 qubits   &           &  2.184  &   2.547  &   +0.363 \\
 7 qubits   &           &  2.285  &   2.602  &   +0.318 \\
 8 qubits   &           &  2.385  &   2.589  &   +0.204 \\
 9 qubits   &           &  2.485  &   2.619  &   +0.134 \\
\hline
\end{tabular*}
\end{center}
\end{table}
We next look at factoring 21 ($3\times 7$).  To do this on a quantum
computer in the same manner as the circuit for factoring 15 shown in
fig.~\ref{fig:15gates} requires a total of 15 qubits, 10 in the larger
register and 5 in the smaller.
For co-prime $x=13$, we find a similar pattern of entanglement to that
shown in fig.~\ref{fig:15pat} for 15 with $x=13$, except that for 21
there is only entanglement between one qubit in the larger register and
two qubits in the lower register.  Similarly, the IQFT step shifts the
entanglement from the top qubit to the bottom qubit in the larger register.
Again, there is no pairwise entanglement, so the three entangled qubits are in
a GHZ type of state \cite{greenberger89a}, i.e., one that can be
rotated into the form $\alpha|000\rangle \pm \beta|111\rangle$.

The larger register is now at the
limit of our computational resources for calculating the full analysis
of the negativity.  Instead of calculating the negativity for all possible
subsets of qubits in a register, we used randomly sampled subsets,
from which we deduce that with high probability
for co-prime $x=13$ there is no entanglement within either register at
any stage of the algorithm.
For other choices of co-prime such as $x=2$ and $x=4$,
there is no entanglement within either register by the end of the
modular exponentiation ($U_x(j)$ gates), but
entanglement is generated within the larger register during the IQFT.
For these co-primes we also find a more 
complex pattern in the entropies of the subsystems:
the entanglement now involves all of the register qubits.
The details for $x=2$ are shown in table \ref{tab:21-2}.
Essentially the entanglement becomes more multipartite:
the average entropy reduces slightly for one and two
qubit subsets, while for larger subsets it increases.
Examination of the entanglement of formation for 
pairs of qubits taken one in each register
shows there is also a significant amount of pairwise
entanglement (average 0.261 per pair before the IQFT) contributing
to the total entanglement in the system.
The average pairwise entanglement of formation between the
registers decreases slightly
(from 0.261 to 0.242) after the IQFT.
This change is possible because entangling gates on the upper register
can convert the pairwise entanglement into something more multipartite
involving more than two of the upper register qubits.
We will discuss what these entanglement patterns can tell us
in the next section after we examine larger examples.

\section{Factoring larger numbers}
\label{sec:entanglement}

In order to examine examples with prime factors larger than
3 or 5, we pushed our numerical studies as far as we could
with this gate model, by analysing semiprimes $32<N<64$ and $64<N<128$,
which require 18 and 21 qubits respectively for the quantum registers.
In these cases, though we cannot easily calculate a full
entanglement analysis, we have calculated
the entropy between one qubit and the rest of the
qubits in both registers, this corresponds to the quantities in the 
third and fourth columns in the top line of table \ref{tab:21-2}.
The difference in the average entropy $\Delta E_1$ before and after the IQFT
(corresponding to the first entry in the fifth column in table \ref{tab:21-2}),
is shown in table \ref{tab:sent1} grouped by the period $r$, as a total
for the whole upper register (so the entry for $N=21$, $r=6$ is $0.624=10\times
0.062$ from table \ref{tab:21-2}).
\begin{table}[!bt]
\begin{center}
\caption{Average decrease in entanglement $-\langle\Delta E_1\rangle$
	between one qubit and the rest during the IQFT step
	(total for upper register, see text for details).}
\label{tab:sent1}
\begin{tabular*}{\textwidth}{@{\extracolsep{\fill}}c|rrrrrrrrr@{}}
\hline \hline
$N$ 		&\multicolumn{9}{c}{$r$ (number of co-primes with this $r$)}\\
$p\times q$ 	&\multicolumn{9}{c}{$-\langle\Delta E_1\rangle$}\\
\hline \hline
$15$ 		&{2 (3)} 		&{4 (4)} 		&{} 	&{} 	&{} 	&{} 	&{}	&{}     &{}\\
$3\times 5$ 	&{0.0} 		&{0.0} 		&{} 	&{} 	&{} 	&{} 	&{}	&{}     &{}\\
\hline \hline
$21$ 		&{2 (3)} 		&3 (2) 		&6 (6) 	&{} 	&{} 	&{} 	&{}	&{}     &{}\\
$3\times 7$ 	&{0.0} 		&0.706 	&0.624	&{} 	&{} 	&{} 	&{}	&{}     &{}\\
\hline \hline
$33$ 		&{2 (3)} 		&5 (4) 		&10 (12) 	&{} 	&{} 	&{} 	&{}	&{}     &{}\\
$3\times 11$ 	&{0.0} 		&0.285 	&0.256		&{} 	&{} 	&{} 	&{}	&{}     &{}\\
\hline
$35$ 		&{2 (3)} 		&3 (2) 		&{4 (4)} 		&6 (6) 		&12 (8)	 	&{} 	&{}	&{}     &{}\\
$5\times 7$ 	&{0.0} 		&0.869		&{0.0}		&0.788 	&0.706 	&{} 	&{}	&{}     &{}\\
\hline
$39$ 		&{2 (3)} 		&3 (2) 		&{4 (4)} 		&6 (6) 		&12 (8)	 	&{} 	&{}	&{}     &{}\\
$3\times 13$ 	&{0.0} 		&0.869		&{0.0}		&0.788 	&0.706 	&{} 	&{}	&{}     &{}\\
\hline
$51$ 		&{2 (3)} 		&{4 (4)} 		&{8 (8)} 		&{16 (16)} 	&{}	 	&{} 	&{}	&{}     &{}\\
$3\times 17$ 	&{0.0} 		&{0.0}		&{0.0}		&{0.0}	 	&{}	 	&{} 	&{}	&{}     &{}\\
\hline
$55$ 		&{2 (3)} 		&{4 (4)} 		&5 (4) 		&10 (12) 	&20 (16)	&{} 	&{}	&{}     &{}\\
$5\times 11$ 	&{0.0} 		&{0.0}		&0.285		&0.256	 	&0.226	 	&{} 	&{}	&{}     &{}\\
\hline
$57$ 		&{2 (3)} 		&3 (2) 		&6 (6) 		&9 (6) 		&18 (18)	&{} 	&{}	&{}     &{}\\
$3\times 19$ 	&{0.0} 		&0.869		&0.788		&0.080	 	&0.071	 	&{} 	&{}	&{}     &{}\\
\hline \hline
$77$ 		&{2 (3)} 		&3 (2) 		&5 (4) 		&6 (6) 		&10 (12) 	&15 (8) 	&30 (24)	&{}     &{}\\
$7\times 11$ 	&{0.0} 		&1.033		&0.343		&0.951	 	&0.314	 	&0.034 	&0.031		&{}     &{}\\
\hline
$91$ 		&{2 (3)} 		&3 (8) 		&{4 (4)} 		&6 (24) 	&12 (32) 	&{}	 	&{}	&{}     &{}	\\
$7\times 13$ 	&{0.0} 		&1.033		&{0.0}		&0.951	 	&0.869	 	&{}		&{}		&{}     &{}\\
\hline
$119$ 		&{2 (3)} 		&3 (2) 		&{4 (4)} 		&6 (6)	 	&{8 (8)}	 	&12 (8)	 	&{16 (16)} & 24 (16)       &48 (32) \\
$7\times 17$ 	&{0.0} 		&1.033		&{0.0}		&0.951	 	&{0.0}	 	&0.869		&{0.0}	 &0.788         &0.706 \\
\hline\hline
\end{tabular*}
\end{center}
\end{table}

The pattern that emerges is that
the closer the period $r$ is to a power of 
2, the smaller the value of $\Delta E_1$.
For $r=2^m$, the IQFT is exact giving $\Delta E_1=0$ in all cases.
\begin{figure}[!tb]
    \begin{center}
	\includegraphics[scale=0.33]{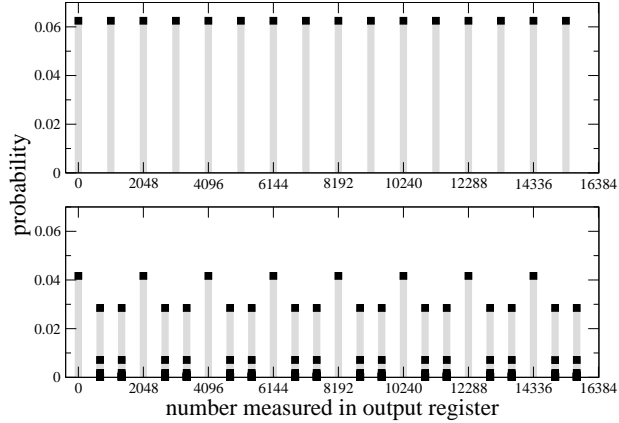}
    \end{center}
    \caption{Distribution of measurement outcomes for factoring 119
	with $x=92$ (upper) and $x=93$ (lower) which have 
	periods $r=16$ and 24 respectively.
	Symbols show the probability of measuring the number on the ordinate
	as the outcome of the algorithm; drop lines are for clarity.
	The upper figure has just 16 peaks, while
	the lower figure shows a significant probability for
	measuring neighbouring numbers to the minor peaks.}
    \label{fig:119drop}
\end{figure}
This can be understood by
looking at the measurement results on the larger register, from which
the period $r$ is calculated.  Figure \ref{fig:119drop} shows the
probability of measuring each possible number $c$ in the larger register
at the end of running the algorithm for two examples: factoring 119 with 
co-prime 92 (period $r=16$), and with co-prime 93 (period $r=24$).
When the period is exactly a power of two, the fraction $c/2^{2n}$
gives the period r exactly, whereas when $r$ is not a power of two,
the peak probability tries to fall between two possible numbers
and thus spreads the wavefunction over several adjacent numbers.
This spread corresponds to increased entanglement in the upper register.
We should note that the case where $r$ is a power of two is rare for
large semiprimes, so the different behaviour for $r=2^m$ does not
help to find a factor.  In fact \cite{pomerance05a},
if $N$ is odd and $r = 2^m$ for all $x$,
it means that the only primes that can divide $N$ are one more than a power
of $2$.  There are only $5$ such odd primes known: 3, 5, 17, 257 and 65537.
It is somewhat weakly conjectured that this is the complete list.  But
even if there are more, factoring such a number is simple: one just
trial divides by numbers one more than a power of 2, more precisely,
of the form $2^{2^m}+1$.  There are approximately $\log\log N$ of them to test,
so this is clearly efficient classically.

\section{Discussion}
\label{sec:discussion}

First, let us summarise what we have found, since there are several steps
to the deductions, necessitated by the limitations of classical computational
power available to us.  In this particular gate model,
the first half of Shor's algorithm, the modular exponentiation, 
generates approximately $\log r$ units of entropy of entanglement
between the two registers \cite{nielsen95a}.
Our simulations suggest that this is the only entanglement at this
point, the entanglement within each register being zero.
By examining the gate sequence within the IQFT we then observed
that the IQFT can only generate entanglement within the upper register,
or, move entanglement around between the upper register qubits.
Our simulations detected both these possibilities: for $N=15$ and $N=21$
with co-prime $x=13$, the entanglement between the registers moves around
the upper register qubits, while for $N=21$ and $x=2$, there is entanglement
generated within the upper register during the IQFT (detected by the
calculating the negativity).
From examining the entropy of subsets of qubits,
we deduced that the overall entanglement becomes more multipartite,
because the entanglement entropy of one and two qubit subsystems decreased,
while that of three to nine qubit subsystems increased.  We then moved
on to larger semiprimes, for which we could only calculate the one qubit
subsystem entropy.  Based on the pattern for $N=21$ just described,
we expect this to \textit{decrease} during the IQFT, corresponding
to \textit{increasing} multipartite entanglement, and this is what
we observed, except where the period $r$ is a power of two, when it
remains equal to zero.

The correlation we observe is between entanglement changes in the
upper register during the IQFT,
and how close the period $r$ is to a power of two.
We can explain this quite easily in terms of the fraction $c/2^{2n}\simeq k/r$
that is being represented in a binary register of size $2n$.
If $r$ is not a power of two, $c$ is trying to fall between two
integers,  and this results in a spread in the
wavefunction in the final state of the quantum register, as shown in
fig.~\ref{fig:119drop}.  Extra terms in the decomposition of the final
state, c.f.~eq.~(\ref{eq:rstate}), correspond to more multipartite
entanglement.  Now suppose we performed the IQFT in
some other base than two -- for example, in base three, perhaps using
a quantum register made up of qutrits (three-state quantum systems)
rather than qubits -- the entanglement pattern would change completely.
The entanglement pattern we have found is thus implementation dependent,
it does not correlate with the success of the algorithm, or the number
being factored.  Nor is any entanglement used up during the course of the
computation.

Our results support the view that entanglement is not used in a
quantitative way to achieve a quantum computation faster than
classical computation.
While entanglement is certainly generated in significant quantities during
pure state quantum computation, this is best understood as a by-product
of exploiting the full Hilbert space for quantum parallelism
\cite{jozsa98a,jozsa02a,greentree03a,blume02a}:
most of Hilbert space consists of
highly entangled states \cite{kendon02a,hayden04a}, so generation of
entanglement during quantum computation is simply unavoidable.

This is to be viewed in contrast with quantum communications tasks,
where maximally entangled pairs of qubits can
perform a specific amount of communication, using up the entangled
pairs in direct proportion to the communication achieved.
We also note that entanglement is used quantitatively
in many practical proposals for implementations of a quantum
computer, notably \cite{raussendorf01a}: this use can be identified
with carrying out communications tasks to move
the quantum data around in the physical qubits.


We thank
Mary Beth Ruskai,
Stephen Parker,
Martin Plenio and
Ben Travaglione
for valuable discussions.
VK was funded by the UK Engineering and Physical Sciences 
Research Council grant number GR/N2507701 (to Sept 2003)
and now by a Royal Society University 
Research Fellowship.  A Royal Society Research Grant provided additional 
computing facilities. 


\small
\bibliography{../bibs/ent,../bibs/qit,../bibs/qrw,../bibs/shor}



\end{document}